\documentclass[conference]{IEEEtran}
\usepackage{graphicx}
\usepackage{booktabs}
\usepackage{caption}
\usepackage{float}
\usepackage{subcaption}
\usepackage[normalem]{ulem}
\usepackage{comment}
\usepackage{mathrsfs}
\usepackage{regexpatch,fancyvrb,xparse}
\usepackage{amsmath}
\usepackage{float}
\usepackage{pdfrender}
\usepackage{scalerel}
\newcommand\vfrac[2]{\ThisStyle{%
  \setbox0=\hbox{$\SavedStyle#1#2$}%
  \setbox2=\hbox{$\SavedStyle X$}%
  \ifdim\ht0>\ht2\setlength{\ht0}{\ht2}\fi%
  #1\mathord{\stretchto{\raisebox{2.3\LMpt}{$\SavedStyle/$}}{\ht0}}#2}}

\usepackage{breqn}
\usepackage{microtype}
\usepackage{caption}
\usepackage{subcaption}
\usepackage{url}
\usepackage{soul}
\usepackage{mathrsfs}
\usepackage{amsmath}
\usepackage{booktabs} 
\usepackage{setspace}
\usepackage{setspace}
\usepackage{subcaption}
\usepackage{verbatim}
\usepackage{courier}
\usepackage{amsmath}
\usepackage{listings}
\usepackage{enumitem}
\usepackage{xspace}
\usepackage{xparse}
\usepackage{enumitem}

\usepackage{hyperref}
\usepackage{xcolor}
\hypersetup{
    colorlinks,
    linkcolor={red!50!black},
    citecolor={blue!50!black},
    urlcolor={blue!80!black}
}

\usepackage[colorinlistoftodos]{todonotes}
\usepackage{xcolor}
\usepackage[most]{tcolorbox}
\usepackage{todonotes}
\usepackage{color,soul}
\usepackage{tikz}

\NewDocumentCommand \T { O{} m } {\ensuremath{\boldsymbol{#1\mathscr{\MakeUppercase{#2}}}}}
\NewDocumentCommand \M { O{} m } {\ensuremath{\bm{#1\mathbf{\MakeUppercase{#2}}}}} 
\NewDocumentCommand \V { O{} m } {\ensuremath{\bm{#1\mathbf{\MakeLowercase{#2}}}}} 

\definecolor{anzheng}{RGB}{255,165,0}
\definecolor{kento}{RGB}{255, 111, 97}
\definecolor{liao}{RGB}{50,125,154}
\definecolor{xin}{RGB}{23,125,54}
\definecolor{tang}{RGB}{255,0,255}
\newcommand{\az}[2][]{#1\textcolor{anzheng}{\textbf{guo:} #2}}

\usepackage[linesnumbered,ruled,vlined]{algorithm2e}
\usepackage[normalem]{ulem}

\usepackage{multirow}
\usepackage{algorithmic}

\makeatletter 
\newcommand{\linebreakand}{%
  \end{@IEEEauthorhalign}
  \hfill\mbox{}\par
  \mbox{}\hfill\begin{@IEEEauthorhalign}
}
\makeatother 

\ifCLASSINFOpdf
\else
\fi
\begin{document}
%
\title{Scrutinizing Variables for Checkpoint Using
Automatic Differentiation}

\author{\IEEEauthorblockN{Xiang Fu, Xin Huang, Wubiao Xu,
 Shiman Meng, Weiping Zhang\\
}
\IEEEauthorblockA{Nanchang Hangkong University\\
\textit{\{fuxiang, xhuang, wxu, smeng, wzhang\}@nchu.edu.cn}}

\linebreakand

\IEEEauthorblockN{
Luanzheng Guo
}
\IEEEauthorblockA{\textit{Pacific Northwest National Laboratory} \\
\textit{lenny.guo@pnnl.gov}}

\and
\IEEEauthorblockN{
Kento Sato
}
\IEEEauthorblockA{\textit{R-CCS, RIKEN } \\
\textit{kento.sato@riken.jp}}
}

\maketitle

\thispagestyle{plain}
\pagestyle{plain}

\begin{abstract}
Checkpoint/Restart (C/R) saves the running state of the programs periodically, which consumes considerable system resources.  
We observe that not every piece of data is involved in the computation in typical HPC applications; such unused data should be excluded from checkpointing for better storage/compute efficiency. 
To find out, 
we propose a systematic approach that leverages automatic differentiation (AD) to scrutinize every element within variables (e.g., arrays) for checkpointing allowing us to identify critical/uncritical elements and eliminate uncritical elements from checkpointing. 
Specifically, we inspect every single element within a variable for checkpointing with an AD tool to determine whether the element has an impact on the application output or not. 
We empirically validate our approach with eight benchmarks from the NAS Parallel Benchmark (NPB) suite.
We successfully visualize critical/uncritical elements/regions within a variable with respect to its impact (yes or no) on the application output. 
We find patterns/distributions of critical/uncritical elements/regions quite interesting and follow the physical formulation/logic of the algorithm. 
The evaluation on NPB benchmarks shows that our approach saves storage for checkpointing by up to 20\%. 
\end{abstract}

\IEEEpeerreviewmaketitle

\section{Introduction}

As High Performance Computing (HPC) systems become increasingly powerful in scale and complexity, failures are observed at a higher frequency at leadership HPC systems~\cite{krasich2009estimate}.
On the other hand, HPC systems are constrained in storage in effect. 
Although the storage capacity is growing rapidly, the datasets are also ever-growing in size and speed~\cite{wang2015performance}.
For example, the digital twin of earth workflow running on Summit at Oak Ridge National Laboratory can generate over 500 TB of data every 15 minutes~\cite{rodd2022doe}.
As a result, storage management remains a critical problem for HPC systems. 

In response to both, we investigate application-level C/R approaches~\cite{hargrove2006berkeley,nicolae2019veloc,sc11:Gomez} and aim to reduce the application states (i.e., variables) necessary for checkpointing. 
Checkpoint/Restart is an essential fault-tolerant approach that stores the running state of the programs periodically, and restarts from the latest stored state in a failure. 
The storage consumption of C/R checkpoints can be very large without careful inspection.
For example, system-level C/R methods (e.g., BLCR~\cite{hargrove2006berkeley}) save all corresponding system state which is significant in storage consumption.
Moreover, at the application level, a variable like a high-dimensional tensor can be costly in storage consumption (e.g., the GPT-3 model which is a tensor that can take 700 GB~\cite{floridi2020gpt} if fully checkpointed). 
However, based on our observation, not every element within the variable participates in computation or can impact the result even if involved in computation.

Our goal is to identify the critical/uncritical elements within the variables for checkpointing. An uncritical element is defined as an element that has no impact on the output; vice versa. 
In particular, we propose an effective approach that uses AD to scrutinize every element within a variable (e.g., arrays) and determine whether it is critical or uncritical. 
This allows uncritical elements to be eliminated from checkpointing for storage efficiency.
Automatic differentiation (AD) is a technology that computes the derivative of the programs, among various existing AD tools we choose Enzyme as the AD tool we use in this study.
We further visualize the distribution of the critical/uncritical elements within each variable for checkpointing.
We also attempt to find the reason why they become critical/uncritical by looking into the source code and algorithm.
We find that in some cases the critical/uncritical elements are determined by the algorithms. For example, 
some elements are written but not read in the following computation because of sampling. 
In other cases, critical/uncritical elements are induced by the source code (code design). For instance, some indexes of an array are declared but not invoked. 
Finally, we evaluate our approach on the NPB benchmark suite and the results show that by eliminating uncritical elements from checkpointing the storage for checkpointing is reduced by an average of 13\% and up to 20\%. 

 

In this paper, our contributions are listed as follows:

\begin{itemize}

    \item 
A novel method that can identify critical elements within variables for checkpointing that have an impact on the execution output without checkpointing the entire variable.

    \item 
Evaluation of the proposed method on the NAS Parallel Benchmarks (NPB) benchmarks.

    \item 
Visualization of the distribution of critical-uncritical elements within variables for checkpointing and investigation into its correspondence to the source code and algorithm.


\end{itemize}
\section{Background}
\subsection{Checkpoint/Restart}

HPC programs typically perform high-intensity complex computations, the large amount of computation and data makes failures more likely to occur. 
Checkpoint/Restart has become a representative of fault-tolerant measures to address this issue through the years. 

Checkpoint/Restart enables programs to store the running state to checkpoint files at a specific interval, and recover from the latest storage once a failure occurs. Users of HPC applications tend to save several versions of checkpoint files to make the data more durable. Due to the different resilience requirement of users, the frequency of checkpointing varies in different HPC applications. 

However, as HPC systems grow in scale, size, and complexity, the occurrence of system failure is observed more frequently. 
To cope with this, the frequency of writing checkpoints has been increasingly higher, which makes the checkpointing overhead higher than ever. 
On the other hand, the amount of data and the number of variables for checkpointing are growing that can also have an significant impact on performance. 
\subsection{Automatic differentiation}

Automatic differentiation (AD)~\cite{} is used for computing the derivative of a function. 
Automatic differentiation is widely employed in many fields, such as discretized non-linear partial differential equation (PDE) solutions \cite{borggaard2000efficient}, stability analysis~\cite{}, uncertainty quantification~\cite{}, silent data corruption (SDC) prediction \cite{menon2018discvar}, etc. AD considers a computer program as a function, which is a combination of a series of basic arithmetic operations that are regarded as primitive functions. AD computes the derivative of the primitive functions to compute the derivative of the output to the program. In this process the chain rule of differential calculus was used, so that the program can compute the derivative of the desired target.
For example, a function looks like:

\begin{equation}
   F = f(y) = f(u(x), v(x))
\end{equation}

Assume that we want to compute the derivative
$\frac{\mathrm{d}f}{\mathrm{d}x}$
. 
There are two main strategies of AD tools: forward mode and reverse mode. The forward mode computes 
$\frac{\mathrm{d}u}{\mathrm{d}x}$
and
$\frac{\mathrm{d}v}{\mathrm{d}x}$
at first, then computes 
$\frac{\mathrm{d}y}{\mathrm{d}u}$
and
$\frac{\mathrm{d}y}{\mathrm{d}v}$
, finally computes
$\frac{\mathrm{d}f}{\mathrm{d}f}$
based on the chain rule.
On the contrary, the reverse mode computes
$\frac{\mathrm{d}f}{\mathrm{d}y}$
and then 
$\frac{\mathrm{d}y}{\mathrm{d}u}$
and
$\frac{\mathrm{d}y}{\mathrm{d}v}$.

\begin{figure}
  \begin{center}
  \includegraphics[width=0.43\textwidth,keepaspectratio]{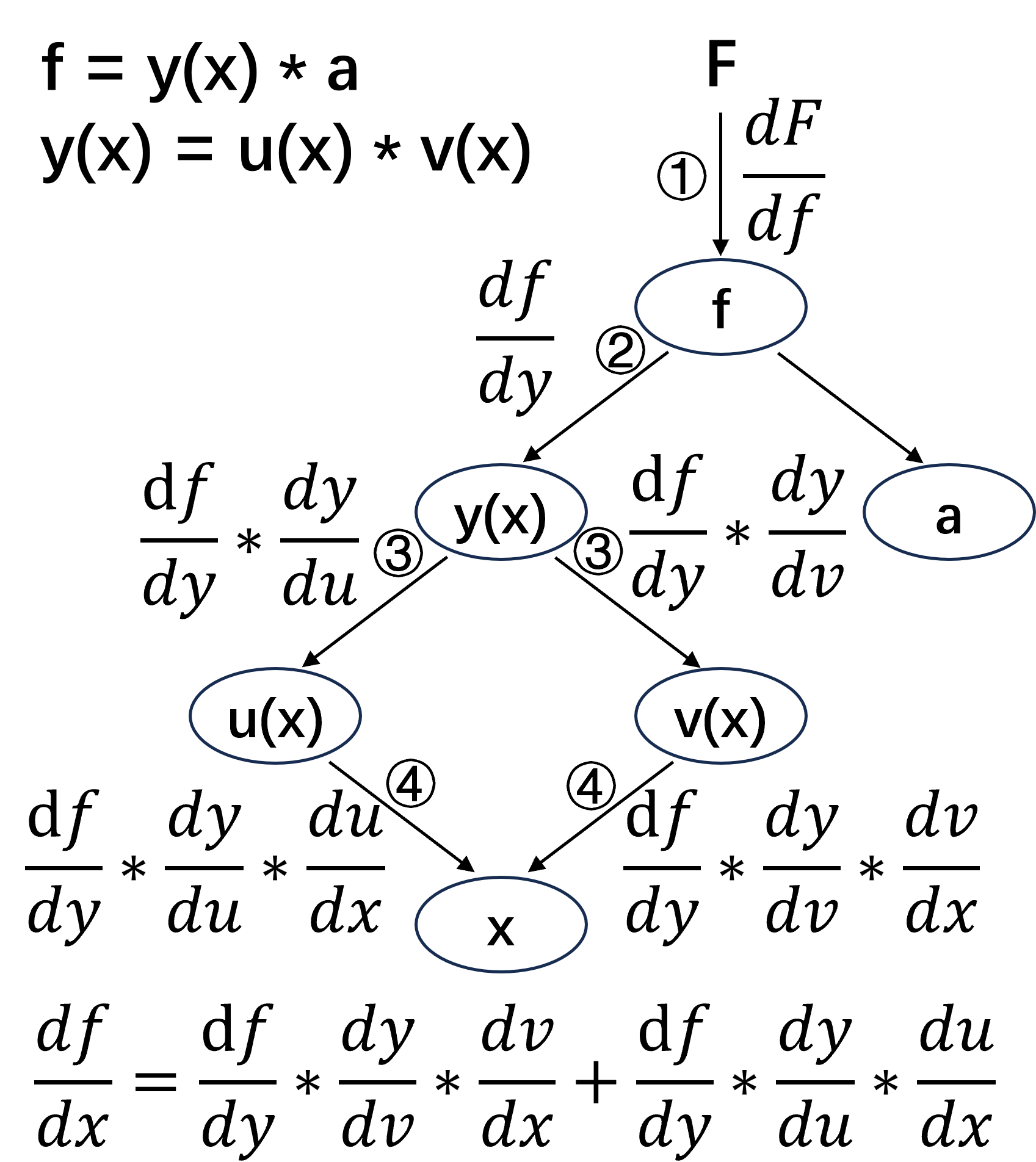}
  \caption{An example of AD workflow. $a$ is a constant.}
  \label{fig: workflow2}
  \end{center}
\end{figure}

We give an example of the reverse mode of AD in Figure \ref{fig: workflow2}.
In reverse mode, AD first sweeps through the forward execution and obtains the information about the program such as variables, branches, and iterations. With the information obtained, AD computes the partial derivative of each operation and calculate the final derivative by the chain rule. 

There are different AD tools like ADIC \cite{bischof1997adic}, Tapenade\cite{hascoet2013tapenade}, OpenAD\cite{utke2008openad},
Enzyme\cite{moses2021reverse}, etc. 
Enzyme\cite{moses2021reverse} is an LLVM compiler-based AD tool that performs reverse mode. Enzyme is a state-of-the-art AD tool that supports various optimizations on different hardware platforms and programming models. 
We thus use it for AD.

\section{Approach}

\subsection{Automatic differentiation}
Our goal is to identify uncritical elements within selected variables necessary for checkpointing that have no impact on the output. 
\emph{A variable is defined as a memory location that is paired with an associated symbolic name. This symbolic name is then referenced in the source code and invoked during execution.} For example, a float-point type array $arr$ declared by $double\ arr[5]$ is variable and $arr[0]$ is an element.
An uncritical element is defined as an element that has zero impact on the output; vice versa.

Specifically, we want to pinpoint, given an element $x$, whether it has impact on the output or not. 
Leveraging AD, we can calculate the derivative of the output with respect to the element $x$. 
Theoretically, if the derivative is 0, the impact of $x$ on the output is 0; otherwise, there is impact on the output. 
We use Enzyme for AD. 
\subsection{Verifying AD result}\label{section: Approach_verifying}

To verify the effectiveness of AD for identifying critical/uncritical elements for checkpointing, we develop a homemade checkpointing library that saves only critical elements to checkpoints. 

We save the location of critical elements in an auxiliary file, which allows us to load individual elements from checkpoints precisely. 
The auxiliary file only records the start and end locations of the region of continuous critical elements. 
With the auxiliary file, the location of critical elements can be determined. 
We only load critical elements while restarting upon a failure.
If the program restarts successfully with expected and correct results generated, we dictate that the AD approach is effective because the critical and uncritical elements are identified correctly. 


\section{Evaluation}

We evaluate our approach with the NPB benchmarks (the C version by Seoul National University).
\emph{First}, we manually determine the variables necessary for checkpointing in each benchmark, followed by our AD approach to identify critical/uncritical elements. 
\emph{Second}, we visualize the distributions of the critical/uncritical elements within each variable necessary for checkpointing. 
\emph{Finally}, we verify the AD results.

\begin{table}[ht]
\begin{center}
\caption {Manually identified variables necessary for checkpointing through trial-and-error. All the variable sizes are determined by the input class of $S$ which is easier to visualize.}
\label{tab:variables_to_checkpoint}
\begin{tabular}{|p{1cm}|p{7cm}|}%
\hline
\textbf{Name}  	& \textbf{Variables and their data structures} \\ \hline \hline
BT  &  double$\quad u[12][13][13][5]$, int$\quad step$	  \\ \hline
SP  & double$\quad u[12][13][13][5]$, int$\quad step$    \\ \hline
MG  & double$\quad u[46480]$, double$\quad r[46480]$, int$\quad it$  \\ \hline
CG  & double$\quad x[1402]$, int$\quad it$  \\ \hline 
LU  & double$\quad u[12][13][13][5]$, double$\quad rho\_i[12][13][13]$, double$\quad qs[12][13][13]$, double$\quad rsd[12][13][13][5]$, int$\quad istep$	 \\ \hline
FT  & dcomple$x\quad y[64][64][65]$, dcomplex$\quad sums[6]$, int$\quad  kt$       \\ \hline
EP  & double$\quad sx$, double$\quad sy$, double$\quad q[10]$, int$\quad k$\\ \hline
IS & int$\quad passed\_verification$, int$\quad key\_array[65536]$, int$\quad bucket\_ptrs[512]$, int$\quad iteration$     \\ \hline
\end{tabular}
\end{center}
\end{table}

\subsection{Variables necessary for checkpointing}
All the variables necessary for checkpointing in the benchmarks are manually determined by \textbf{trial-and-error} listed in Table \ref{tab:variables_to_checkpoint}. 
We consider all of them in our study.
The variables are described as follows:


BT and SP: $u$ is the function to solve for nonlinear partial differential equations (PDEs); $step$ is the main loop index.
\emph{The main loop is the outermost loop within the main function.} 

MG: $u$ is the solution to the three-dimensional discrete Poisson equation; $r$ is the residual of the equation; $it$ is the main loop index.

CG: $x$ is the input vector of the linear system of equations; $it$ is the main loop index.

LU: $u$ is the function to solve for nonlinear partial differential equations (PDEs); $rho\_i$ is the relaxation factor in the Symmetric Successive Over-Relaxation method; $qs$ is a variable for computing the flux differences; $rsd$ is a variable for computing the final residual.

FT: $y$ is the output signal of Fast Fourier Transform at frequency domain; Array $sums$ aggregates the sums computed from all iterations; Variables $y$ and $sums$ are of custom data type $dcomplex$, containing two attributes, $real$ of \texttt{double} and $imag$ of \texttt{double}.
$kt$ is the main loop index.

EP: $sx$ and $sy$ are the sums of independent Gaussian deviates at the $X$ and $Y$ dimensions respectively; $q$ is the number of pairs of coordinates at $X$ and $Y$; $k$ is the main loop index.

IS: $passed\_verification$ is the verification counter; $iteration$ is the main loop index; $key\_array$ is the array that stores the keys of bucket sort; $bucket\_ptrs$ is the pointer of the `bucket’ in bucket sort.


\subsection{Automatic differentiation analysis}
\textbf{BT \cite{bailey1993parallel}}:
BT is one of the NPB benchmarks that performs Block Tri-diagonal solver to solve the three sets of equations. 
For BT, there are two variables necessary for checkpointing, a four-dimensional array $u$ and an integer $step$.
$step$ is the index of the main loop. 
$step$ is a scalar that has an impact on the output as it is necessary for checkpointing. 
Its impact is obvious as the index variable of a for-loop. 
$u$ contains 10,140 elements with the input class of $S$. 
We calculate the derivative at each element of $u$ by AD to pinpoint their impact on the output. 
There are 8640 critical elements (there is impact) and 1500 uncritical elements (no impact).
The uncritical elements are accounted for 14.8\% of the total elements within $u$.

\begin{figure}
  \begin{center}
  \includegraphics[width=0.45\textwidth,keepaspectratio]{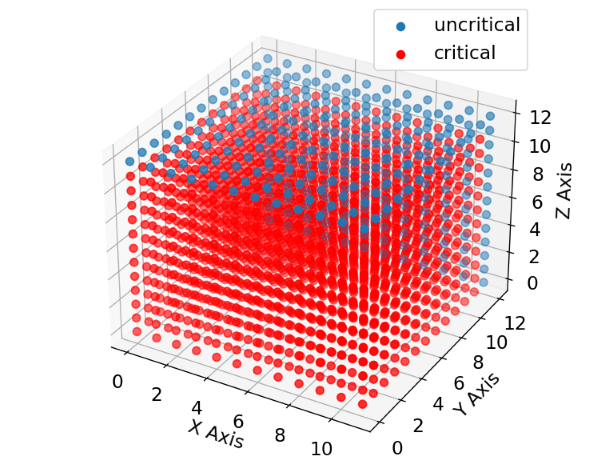}
  \caption{A typical critical-uncritical distribution in NPB benchmarks(red: critical, blue: uncritical). Variables following this distribution: $u(BT)$, $u(SP)$, $u[x][y][z][0](LU)$, $u[x][y][z][1](LU)$, $u[x][y][z][2](LU)$, $u[x][y][z][3](LU)$, $rho\_i(LU)$, $qs(LU)$, $rsd(LU)$}
  \label{fig: typical pattern of 3d/4d variables}
  \end{center}
\end{figure}

To further understand the distribution of critical and uncritical elements, we aim to visualize the distribution of critical-uncritical elements. 
This is very challenging as $u$ is a $12\times13\times13\times5$ four-dimensional array.
Fortunately, we find that $u$ can be decomposed into five three-dimensional arrays of $12\times13\times13$.  
We find that all five three-dimensional arrays share the same critical-uncritical distribution pattern. 
We show one of the three-dimensional arrays in Figure \ref{fig: typical pattern of 3d/4d variables}.
The critical-uncritical distribution is quite interesting, in which uncritical elements are represented in blue and critical elements are represented in red; the uncritical elements are distributed on the two surfaces of the cube at $y=12$ and $z=12$; the remaining elements are critical.

\begin{figure}
  \begin{center}
  \includegraphics[width=0.45\textwidth,keepaspectratio]{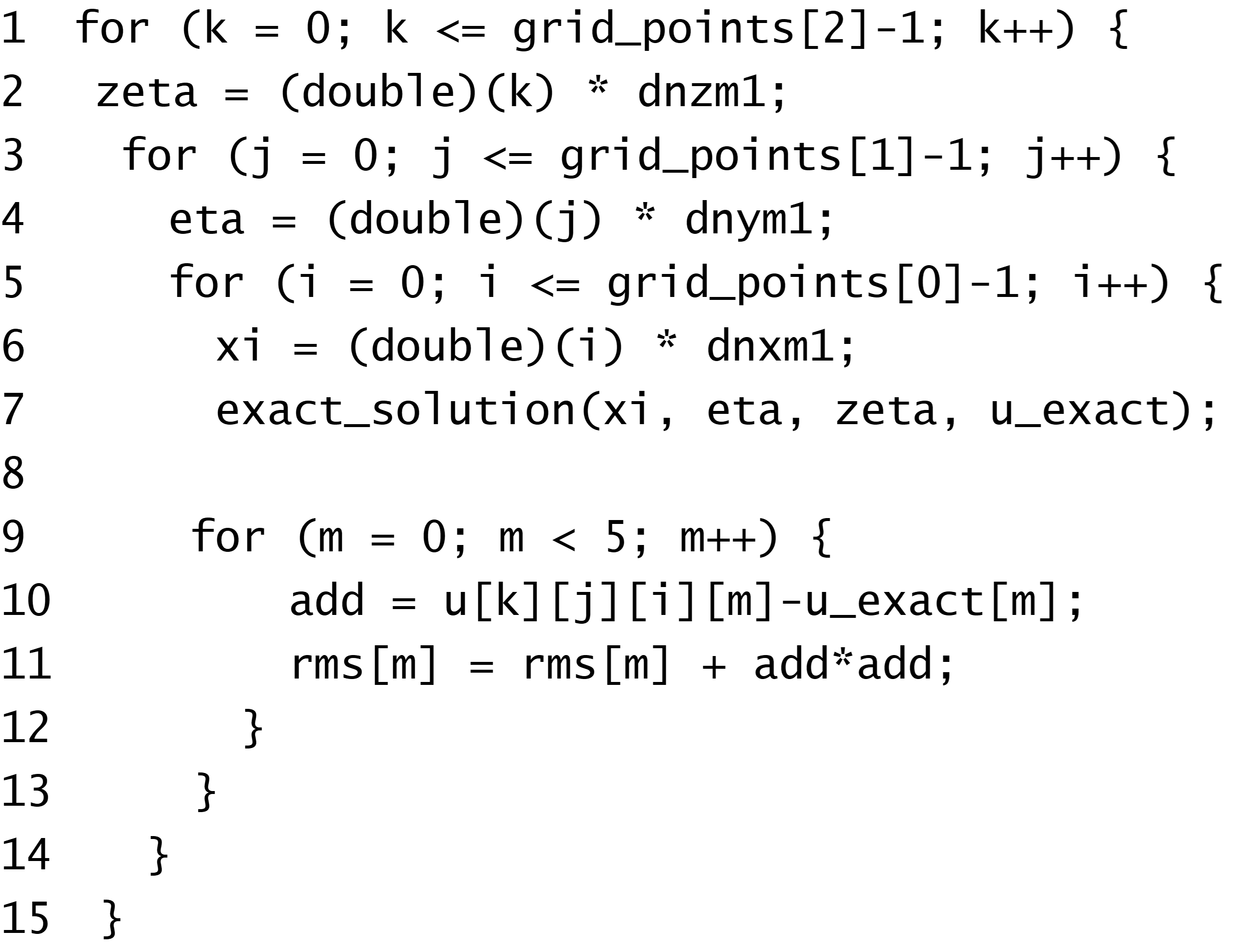}
  \caption{Source code of the function $error\_norm$ in $BT$}
  \label{fig: erro_norm}
  \end{center}
\end{figure}

We attempt to understand the underlying logic of the distribution by making connections to the source code and algorithm. 
After digging into the source code, we find that the $error\_norm$ function is the one that creates the specific critical-uncritical distribution in $u$.
The $error\_norm$ function can be found at Line 41 in the \texttt{error.c} file of the BT source code. 
We provide the part of the source code of $error\_norm$ that operates on $u$ in Figure~\ref{fig: erro_norm}, where $u$ is used at Line 10. 
We observe that the range variables, $grid\_points[0]$ (see Line 5), $grid\_points[1]$ (see Line 3), and $grid\_points[2]$ (see Line 1), all have a value of 12.
Given fixed $m$, recalling that $u$ is $u[12][13][13][5]$ while the access range is from zero to 11 for $k$, $j$, and $i$, we observe that the elements are not used at $j=12$ and $i=12$, mapping to axes Y and Z in Figure~\ref{fig: typical pattern of 3d/4d variables}. 
Therefore, the elements at $y=12$ and $z=12$ are uncritical because they did not participate in computation. 

This is an interesting finding when uncritical elements are caused by imperfect coding. 
Algorithmically, $u$ ought to be checkpointed fully.

\textbf{SP \cite{bailey1993parallel}}:
$SP$ is a Scalar Pentadiagonal solver similar to $BT$. 
The analysis process is similar to $BT$ because of the similar code structure to $BT$. The two variables necessary for checkpointing in $SP$ are the same as $BT$, a four-dimensional array u, and an integer $step$. 
We find the exactly same critical-uncritical distribution in $u$ as we found in $u$ in BT.
Again, the $error\_norm$ function operates on $u$ and creates the particular critical-uncritical distribution in $u$.
$SP$ invokes the same function $error\_norm$ at Line 41 in \texttt{error.c}, which is the exactly same as is invoked in $BT$. 
$step$ is the index of the main loop that is needed for checkpointing.


\textbf{MG \cite{bailey1993parallel}}:
The MultiGrid algorithm employs the V-cycle multigrid method to efficiently solve a three-dimensional discrete Poisson equation. Our method is evaluated on $MG$ of NPB benchmarks with the input size $S$. 
The variables necessary for checkpointing in this program are integer $it$, array $u$ of 46480 elements, and array $r$ of 46480 elements.
$it$, the index of the main loop, is critical for checkpointing.

\begin{figure}
  \begin{center}
  \includegraphics[width=0.45\textwidth,keepaspectratio]{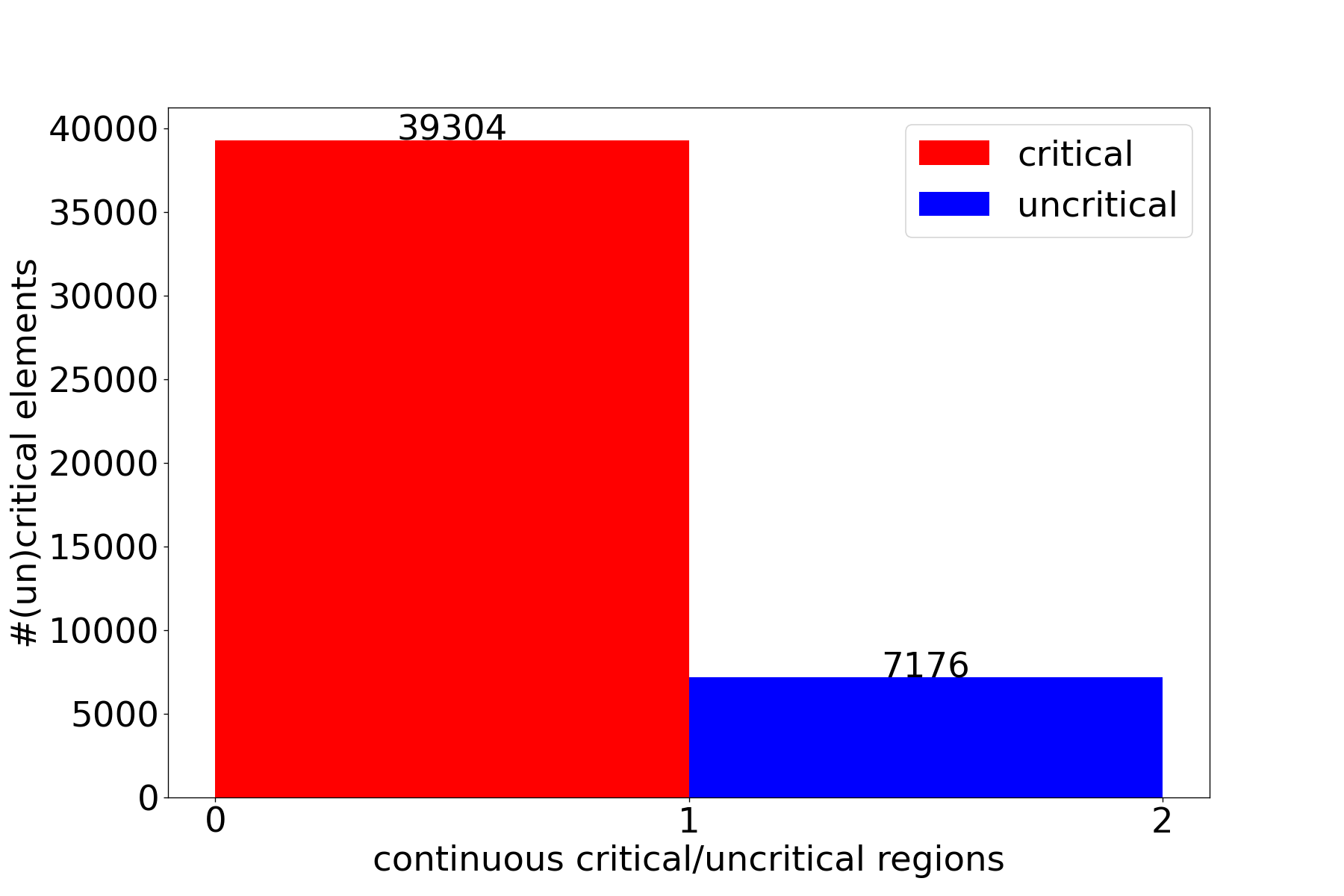}
  \caption{Critical-uncritical distribution of array $u$ in MG}
  \label{fig: MG_u}
  \end{center}
\end{figure}

\begin{figure}
  \begin{center}
  \includegraphics[width=0.45\textwidth,keepaspectratio]{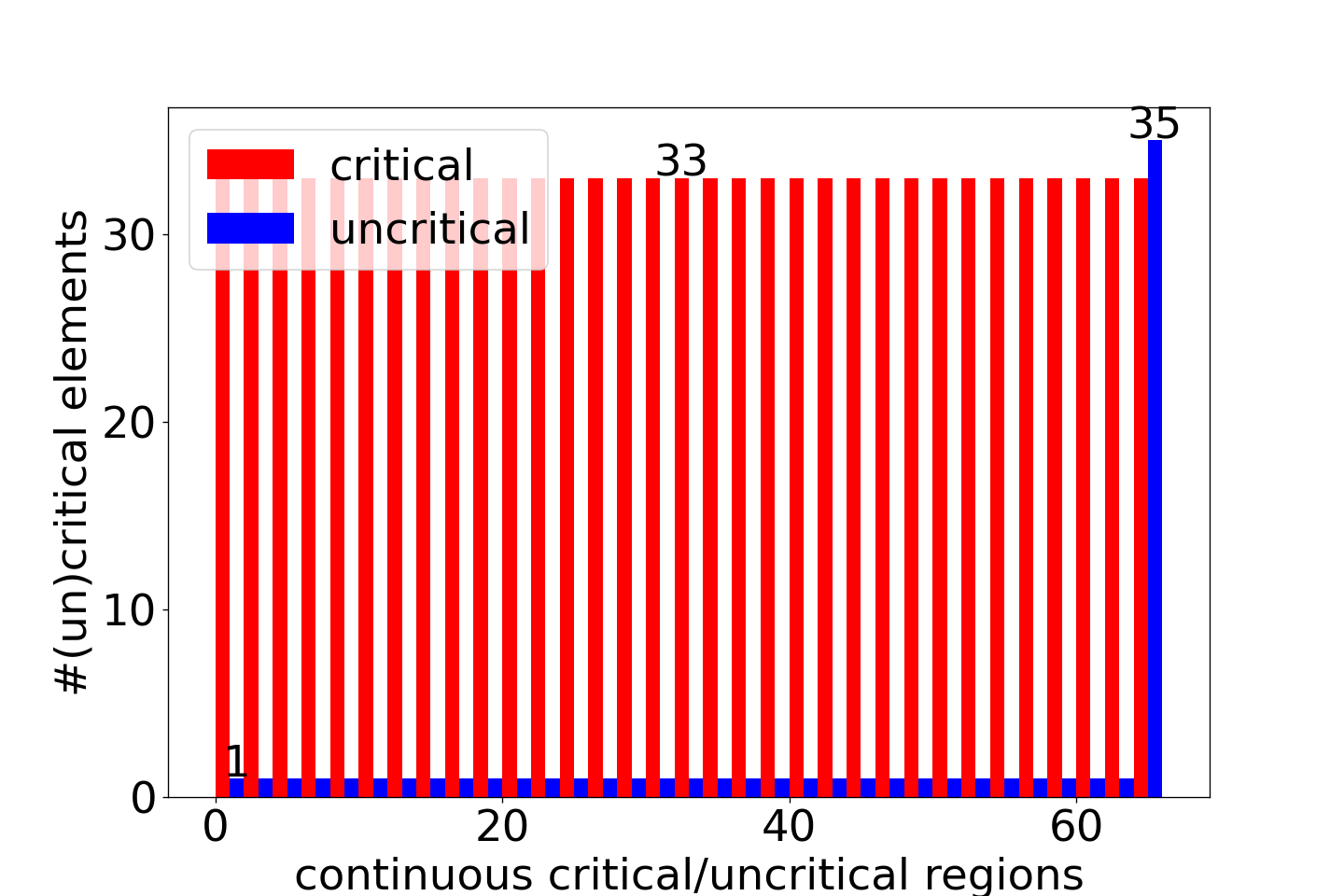}
  \caption{Critical-uncritical distribution of of array $r$ in MG}
  \label{fig: MG_r}
  \end{center}
\end{figure}

We find 7176 uncritical elements in $u$ and 10479 uncritical elements in $r$, respectively, accounting for 15.3\% and 22.4\% of all the elements. 
We find that, the variable $u$ is transformed into a $34\times34\times34$ three-dimensional array in $MG$ to be used in computation.
In effect, there are only $34\times34\times34$ elements of $u$ participating in the computation.
We visualize the critical-uncritical distribution of $u$ in Figure \ref{fig: MG_u},
which shows that there are 39304 ($34\times34\times34$) continuous critical elements, followed by 7176 continuous uncritical ones within array $u$. 
However, $r$ has a more complex pattern of critical-uncritical distribution as the invoked elements are triggered by an integer array $ir$.
Figure \ref{fig: MG_r} shows a repetitive pattern \textbf{as part of} the critical-uncritical distribution of $r$. 


\begin{figure}
  \begin{center}  \includegraphics[width=0.45\textwidth,keepaspectratio]{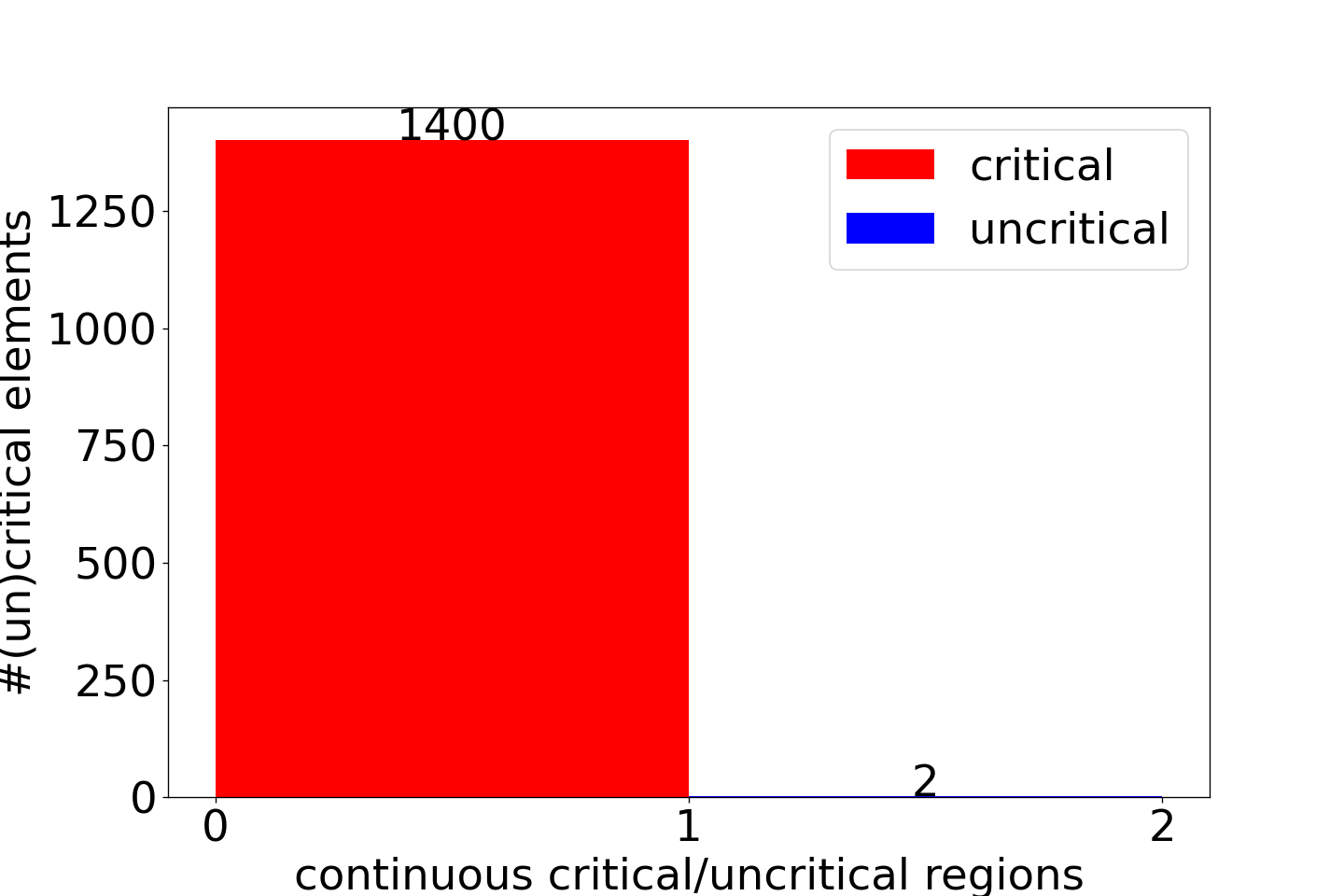}
  \caption{Critical-uncritical distribution of array $x$ in CG}
  \label{fig: CG_x}
  \end{center}
\end{figure}

\textbf{CG \cite{bailey1993parallel}}:
The Conjugate Gradient method utilizes the conjugate gradient and inverse iteration methods as a solution to linear equations.
There are two variables necessary for checkpointing in CG: an integer $it$ that is the index of the main loop, and an array $x$ of length 1402.
Figure \ref{fig: CG_x} shows the distribution of critical-uncritical regions of $x$ in CG. The first 1400 continuous elements are critical, and the remaining 2 elements are uncritical. 
$x$ is first read as an input and then updated (written) within the main loop. 
After delving into the source code, we find that $x$ has a size of $NA+2$, in which $NA$ is a macro that has the value of 1400 for the $S$ class, and only the first $NA$ elements participate in computation. 
$it$ is required for checkpointing.

\textbf{LU \cite{bailey1993parallel}}:
LU is the Lower-Upper Symmetric Gauss-Seidel solver, which is a numerical method to solve linear systems of equations.
There are five array variables necessary for checkpointing in LU, which are $u$, $rho\_i$, $qs$, $rsd$, and $istep$.
$istep$ is the index of the main loop necessary for checkpointing.

\begin{figure}
  \begin{center}  \includegraphics[width=0.45\textwidth,keepaspectratio]{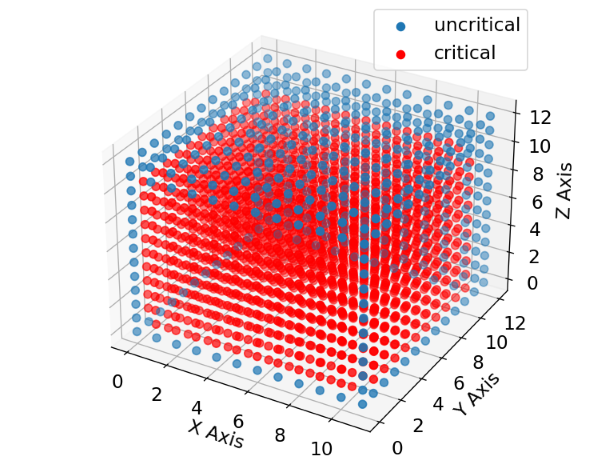}
  \caption{Critical-uncritical distribution of $u[x][y][z][4]$ in $LU$ (red: critical, blue: uncritical) }
  \label{fig: LU_u}
  \end{center}
\end{figure}

\textbf{$u$}: There are 10140 elements in array $u$, in which 1628 elements are uncritical, accounting for 16\% of all elements. 
The variable $u$ here is slightly different from $u$ in $BT$ and $SP$, although they are in the same data type and size (i.e., double $u$[12][13][13][5]).
There are five $12\times13\times13$ 3D arrays in $u$.
We find that each of the first four $12\times13\times13$ 3D arrays follows the critical-uncritical distribution in Figure~\ref{fig: typical pattern of 3d/4d variables} exactly.
However, the fifth, $u[x][y][z][4]$, follows a different critical-uncritical distribution shown in Figure~\ref{fig: LU_u}.

After delving into the source code, we find that $u[x][y][z][4]$ participates in multiple discontinuous computations unlike $u[x][y][z][0-3]$ that is utilized in separate computations akin to Figure~\ref{fig: erro_norm}.
We find that the components of $u[x][y][z][4]$ used in multiple discontinuous computations are $u[1-10][1-10][0-11][4]$, $u[1-10][0-11][1-10][4]$, and $u[0-11][1-10][1-10][4]$, which constitute the critical (red) area in Figure~\ref{fig: LU_u}.
We also find that there are 128 more uncritical elements (on the edges) not participating in computation compared with the critical-uncritical distribution in Figure \ref{fig: typical pattern of 3d/4d variables}.



\textbf{$rho\_i$ and $qs$}:
We find that there are 300 uncritical out of 2028  elements in $rho\_i$.
Particularly, there are $12\times12\times12$ out of $12\times13\times13$ elements participating in computation, which leads to the same critical-uncritical distribution in Figure \ref{fig: typical pattern of 3d/4d variables}.
It is the same case for $qs$.


\textbf{$rsd$}: 
$rsd$ is exactly the same as $u$ in BT: they are in the same size and participate in the same computation that results in the same critical-uncritical distribution (Figure~\ref{fig: typical pattern of 3d/4d variables}).

\textbf{FT \cite{bailey1993parallel}}:
FT performs a 3D Fast Fourier Transform on a grid of three-dimensional points.
There are three variables necessary for checkpointing in $FT$: a $64\times64\times65$ three-dimensional array $y$, containing 266,240 elements, and each element is a custom data structure $dcomplex$ with two attributes $imag$ and $real$ of floating point type; a $dcomplex$ array $sums$ of length 6; and an integer $kt$. 
$kt$ is the index of the main loop required for checkpointing.

\begin{figure}
  \begin{center}  \includegraphics[width=0.45\textwidth,keepaspectratio]{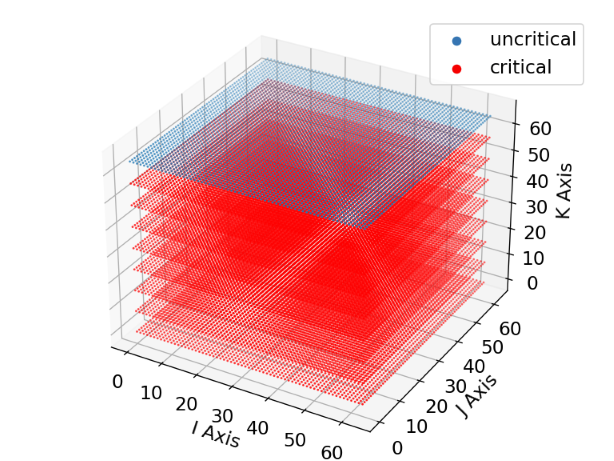}
  \caption{Critical-uncritical distribution of $y$ in $FT$}
  \label{fig: FT_y}
  \end{center}
\end{figure}


$y$: we find 4096 uncritical of all 266240 elements.
Figure \ref{fig: FT_y} shows the critical-uncritical distribution,
in which only the top layer (in blue) at $k=64$ does not participate (not used) in computation. 
This is due to imperfect coding, which can be avoided not only for code safety but also for efficient usage of memory and storage and also for checkpointing efficiency.

$sums$: $sums$ stores the result computed at each iteration of the main loop. 
Therefore, it is critical to write it to storage to avoid loss of computed results upon a failure.



\textbf{EP \cite{bailey1993parallel}}:
The embarrassingly parallel (EP) generates a pair of random numbers that follow a normal distribution.
The variables necessary for checkpointing are floating-point $sx$ and $sy$, array $q$, and integer $k$.
Both $sx$ and $sy$ are write-after-read which is necessary for checkpointing.
$k$ is the index of the main loop that is required for checkpointing.
$q$ stores the output at each iteration of the main loop, so we must write it to storage to avoid recomputation upon a failure. 

\textbf{IS \cite{bailey1993parallel}}:
$IS$ is a bucket sorting method specifically designed for sorting small integers.
The variables necessary for checkpointing in $IS$ are integer $passed\_verification$, integer $iteration$, integer arrays $key\_array$, and $bucket\_ptrs$.

Again, $passed\_verification$ is write-after-read, which is necessary for checkpointing. $iteration$ is the index of the main loop that is a critical variable for checkpointing.
Like $iteration$, $key\_array$ and $bucket\_ptrs$ store the indexes for other arrays which makes them critical for checkpointing.

We summarize the number of uncritical elements along with the respective percentage they represent in relation to the total elements for all variables necessary for checkpointing in Table~\ref{Tab: uncritical number}.

\subsection{Verifying AD results}

To verify that the uncritical elements detected by AD are not needed and the critical elements are critical for checkpointing, we implement a homemade checkpointing library that writes only critical elements to checkpoint files.
All NPB benchmarks have their own verification phase, which uses a margin of error to determine if the computation is successful or failed. 
We rely on their verification to determine the AD result's correctness. 
In principle, the uncritical elements should not impact the computation correctness even if their values are altered by system failures. 
On the other hand, the critical elements must impact the execution output, and the verification is expected to fail if they cannot recover from failures. 
It turned out that, all benchmarks restarted successfully and passed the verification upon only checkpointing the critical elements. 
This demonstrates the effectiveness of the AD analysis for scrutinizing variables for checkpointing. 
\subsection{Storage for checkpointing}
We show the comparison of checkpointing storage before and after eliminating uncritical elements in Table \ref{tab: storage table}. As it shows, the storage saved is consistent with the uncritical rate in Table \ref{Tab: uncritical number}.

\begin{table}[]
    \centering
    \begin{tabular}{cccc}
    \toprule[2pt]
         Benchmark(variable) &Uncritical& Total & Uncritical rate\\
    \midrule[1pt]
         BT(u) & 1500 & 10140 & 14.8\%\\
         SP(u) & 1500 & 10140 & 14.8\%\\
         MG(u) & 7176 & 46480 & 15.4\%\\
         MG(r) & 10543 & 46480 & 22.7\%\\
         CG(x) & 2 & 1402 & 0.1\%\\
         LU(qs) & 300 & 2028 & 14.8\%\\
         LU(rsd) & 300 & 2028 & 14.8\%\\
         LU(rho\_i) & 1500 & 10140 & 14.8\%\\
         LU(u) & 1628 & 10140 & 16.0\%\\
         FT(y) & 4096 & 266240 & 1.5\%\\
    \bottomrule[2pt]
    \end{tabular}
    \caption{Number of uncritical elements}
    \label{Tab: uncritical number}
\end{table}

\begin{table}[]
    \centering
    \begin{tabular}{cccc}
    \toprule[2pt]
         Benchmark & Original & Optimized & Storage saved\\
    \midrule[1pt]
         BT & 79.4kb & 67.7kb & 14.8\%\\
         SP & 79.4kb & 67.7kb &14.8\%\\
         MG & 727kb & 588kb &19.1\%\\
         CG & 10.9kb & 10.9kb &0.1\%\\
         LU & 191kb & 161kb &15.7\%\\
         FT & 4161kb & 4097kb &1\%\\
    \bottomrule[2pt]
    \end{tabular}  
        \caption{Checkpointing storage}
    \label{tab: storage table}
\end{table}

\section{Discussion}
The uncritical elements for checkpointing we find in the NAS Parallel Benchmark suite are all not involved in the computation, which is caused by programming defects. Ideally, AD analysis is not needed in the future if the programming defects can be avoided.
We hope it is possible to find the uncritical elements of variables necessary for checkpointing by algorithmic analysis rather than AD analysis.
\section{Related work}

\textbf{Checkpoint/Restart.} 
Various C/R libraries have been created for various purposes through the years.
The main idea of C/R is to save the running state of programs in persistent storage and restart the program from the latest checkpoint files once the failures occur.
Moody et al. \cite{Moody:Scalable_sc_03_reduce}developed a scalable checkpoint/restart (SCR) library, a multi-level checkpoint system that can save checkpoints to the computing node's RAM, Flash, disk, and parallel file system.
Hargrove et al. \cite{hargrove2006berkeley}created the Berkeley Lab Checkpoint/Restart (BLCR) library, which provides system-level checkpointing for the Linux kernel.
Gholami et al. \cite{gholami2021combining}combined XOR and partner checkpointing to develop a stable and efficient C/R approach.
Vasavada et al. \cite{vasavada2010innovative}proposed a page-based incremental checkpoint approach by which memory writes are tracked by trapping dirty pages to be saved. 
The Fault Tolerant Interface (FTI) \cite{sc11:Gomez}is a three-level checkpoint scheme with a topology-aware Reed-Solomon encoding integration.

However, the existing checkpoint approaches can be further optimized to some extent. To the best of our knowledge, there is no such application-level optimization that can analyze the semantics of programs to optimize the checkpoint overheads. In this mind, we propose a brand new method for identifying the critical elements in variables for checkpoint based on the semantics of the program.

\textbf{Automatic differentiation.} 
Paszke et al.\cite{paszke2017automatic} provide a library that aims at rapid research on various machine-learning models, an automatic differentiation module of Pytorch.
Minkov et al. \cite{minkov2020inverse}use an automatic differentiation library to optimize the ultrasmall cavity’s quality factor and the dispersion of a photonic crystal waveguide.
Krieken et al. \cite{krieken2021storchastic}propose a stochastic automatic differentiation framework that minimizes the gradient estimates' variance.
Virmaux et al. \cite{virmaux2018lipschitz}provide an algorithm working with automatic differentiation, which benefits the computations for extending and improving estimation methods.
Crooks \cite{crooks2018performance}provides a study on the optimization of quantum circuits using automatic differentiation for the maximum cut problem.

Although automatic differentiation is applied in many fields of research, there is no existing research on optimizing Checkpoint/Restart as far as we know. We first use AD to compute the impact of elements with the variables necessary for checkpointing on an application level.
\section{Conclusion and future work}
In general, our method creatively applies the perspective of checkpoint variables to the elements within. This approach can identify the portion of elements that are not critical within variables necessary for checkpointing demonstrated with NPB benchmarks. 
We use the AD tool to find these elements because they are possibly redundant elements for checkpointing. 
The result is that all the elements that are considered to have no impact on the output are not involved in the computation.
We visualize the critical/uncritical distribution of elements within variables necessary for checkpointing. 
With visualizations like this, we develop in-depth insights into data necessary for checkpointing how the critical and uncritical elements are distributed.
Then we modify a homemade checkpoint library to verify our conclusions. By decreasing the number of elements for checkpointing, we successfully restart the programs with a correct output while the rate of uncritical elements in all reduces the storage occupation of checkpoint files.



Our work is not only for HPC application checkpoint but also potentially benefits to accelerate applications by using lower precision for uncritical or even those elements that are of very low impact in the future. 


\bibliographystyle{IEEEtran}
\bibliography{ref} 

\end{document}